\documentclass[aps,prl,superscriptaddress,twocolumn,showpacs]{revtex4-1}
\usepackage[pdftex]{graphicx}
\usepackage{xcolor}
\usepackage{placeins}
\usepackage{amsmath}

\begin{document}


\title{ Coupling between Rydberg states and Landau levels of electrons trapped on liquid helium }


\author{K.M. Yunusova}
\affiliation{LPS, Univ. Paris-Sud, CNRS, UMR 8502, F-91405, Orsay, France}
\affiliation{Institute of Physics, Kazan Federal University, Kazan, 420008, Russian Federation}

\author{ D. Konstantinov}
\affiliation{Okinawa Institute of Science and Technology (OIST) Graduate University, Onna, Okinawa 904-0412, Japan}

\author{ H. Bouchiat}
\affiliation{LPS, Univ. Paris-Sud, CNRS, UMR 8502, F-91405, Orsay, France}

\author{A.D. Chepelianskii}
\affiliation{LPS, Univ. Paris-Sud, CNRS, UMR 8502, F-91405, Orsay, France}


\date{\today}

\begin{abstract}
We investigate the coupling between Rydberg states of electrons trapped on a liquid Helium surface and Landau levels induced by a perpendicular magnetic field. We show that this realises a prototype quantum system equivalent to an atom in a cavity, where their coupling strength can be tuned by a parallel magnetic field. We determine experimentally the renormalisation of the atomic transition energies induced by the coupling to the cavity, which can be seen as an analogue of the Lamb shift. When the coupling is sufficiently strong the transition between the ground and first excited Rydberg states splits into two resonances corresponding to dressed states with vacuum and one photon in the cavity. Our results are in quantitative agreement with the energy shifts predicted by the effective atom in a cavity model where all parameters are known with high accuracy.
\end{abstract}


\maketitle

The realization of high purity two dimensional electron systems (2DES) has lead to the discovery of fundamental new states in condensed matter physics like integer and fractional quantum Hall effects \cite{halleffect1,halleffect2,halleffect3,halleffect4,halleffect5}, as well as to the more recent discovery of two dimensional topological insulators \cite{top_insul}. Electrons on liquid helium were one of the first historical realisations of the 2DES \cite{twodes1,twodes2,twodes3,twodes4}. This system is formed due to the attractive interaction between electrons and their image charge inside liquid helium, it achieves an exceptional purity and still gives the best known electronic mobilities for a 2DES \cite{mobility}. Electrons on helium enabled the first observation of Wigner crystallization \cite{wignercryst1,fisher1979phonon}, edge magnetoplasmons \cite{first_magnetoplasmons} and other exciting many-electron phenomena \cite{dyk_khazan,dyk_lea_foz_frost,konst_dyk_lea_monar_kono,ikeg_aki_rees_kono,rees_niyaz_lin_kono}. Considerable efforts were also devoted to study the interaction between electrons on helium and millimeter-wave photons aiming for applications in quantum computing \cite{quantumcomp,SchusterPRL}. This research direction recently revealed a rich nonequilibrium physics, showing microwave induced oscillations (MIRO) \cite{miro1,miro2,miro4,miro3,miro5}, zero-resistance states \cite{zrs1,zrs2,zrs3} and incompressible electronic behavior \cite{konst_alexei_kono,alexei2015,monarkha2016} under excitation by millimeter-wave photons.
In the present Letter we show that electrons on helium also allow to realize a model system for an atom interacting with an oscillator (cavity) and to explore its physical properties directly controlling their coupling with a parallel magnetic field. Such systems have been embodied for example in atomic physics \cite{haroche_nobel} and quantum optics as well as with superconducting circuits  \cite{circuit_qed,spin_cavity}. \textcolor{black}{We demonstrate that, for weakly coupled electrons, the quantum electro-dynamics (QED) Hamiltonian reproduces quantitatively the spectroscopic properties of our system. This opens a doorway to study quantum phenomena in an ensemble of interacting atoms in a cavity systems by tuning the strength of electron-electron interactions.}

Before presenting our experimental results, we describe the derivation of the QED Hamiltonian for electrons on helium. The electric field of an electron polarizes the liquid helium around it and creates an image charge that attracts it towards the helium surface; a steep electron-volt high energy barrier prevents it from penetrating inside the liquid helium. The interaction with the image charge gives rise to a one-dimensional Coulomb potential which leads to the quantization of the vertical motion and to the formation of a Rydberg series of bound states for a one-dimensional hydrogen-like atom. This series will play the role of the atomic degree of freedom in our QED model. A pressing perpendicular static electric field $E_{\perp}$ is also present in the experiments, it allows to shift the Rydberg levels through linear Stark effect \cite{lea}. The spectroscopic positions of the Rydberg states is well described by a one-dimensional Schr\"odinger equation for vertical motion:

\begin{equation}
	\label{atom_energy}
	\mathcal{H}_{a} =-\frac{\hbar^{2}}{2m}\frac{\partial^{2}}{\partial z^{2}}+V_{a}(z) = \sum_\alpha \varepsilon_{\alpha}\left|\alpha\right\rangle \left\langle \alpha\right|
\end{equation}

\noindent where we introduced $z$ the vertical distance of the electrons to the helium surface, the eigenstates for the vertical motion $|\alpha\rangle$ and their eigenenergies $\epsilon_\alpha$. Above the helium surface, for $z > 0$, the confinement potential $V_a(z)$ is the sum of the interaction with the image charge and with the perpendicular electric field $V^0_{a}(z) = -\Lambda/z - eE_{\perp}z$ with $\Lambda = \frac{e^{2}\left(\varepsilon-1\right)}{16 \pi \left(\varepsilon+1\right)}$ and where $\varepsilon$ is liquid helium's dielectric constant. On the energy scale of the bound states ($ \sim 7$~K) we can set $V_a(z) = \infty$ inside liquid helium for $z<0$. We introduced a subscript $V^0_{a}(z)$ to the potential since we will show later that $V_a(z)$ is renormalized when an in plane magnetic field is present. For usual pressing electric fields $E_{\perp} \sim 2{\;\rm V\;mm^{-1}}$ the main contribution to the confinement potential for the lowest eigenstates comes from the interaction with the image charge.

In addition to their vertical motion, electrons on helium move horizontally as free particles - electrons with their bare electronic mass $m$. A perpendicular magnetic field applied to 2DES induces the Landau quantization of horizontal motion and the formation of equidistant Landau levels, the Hamiltonian for horizontal motion (up to a constant) then becomes $\mathcal{H}_{l} = \hbar \omega_c {\hat a}^+ {\hat a}$ where $\omega_c = e B_z/m$ is the cyclotron frequency. \textcolor{black}{This term has the same form as the Hamiltonian of a resonant cavity in QED, the Landau level index then plays the role of the number of light quanta in the cavity}. With only a perpendicular magnetic field and in the limit of weak electron-electron interaction the Landau levels and Rydberg states are not coupled. A tunable coupling can be introduced by applying an in plane magnetic field \cite{Main,Shepelyansky}. Indeed a magnetic field applied in the $y$ direction will tend to turn a vertical velocity towards the $x$ direction due to cyclotron motion along the $y$ axis induced by the parallel field. \textcolor{black}{ This coupling has been investigated in double quantum wells in a regime with many occupied Landau levels  \cite{pfeiffer91,vitkaov16a,vitkaov16b}, in this letter we focus instead on the limit where only the lowest Landau level is occupied}. The quantitative form of the interaction induced by the in plane field can be obtained as follows. We write the total Hamiltonian $\mathcal{H}= \frac{\left({\mathbf{p}}-e{\mathbf{A}}\right)^{2}}{2m}+V_a^0(z)$, using the Landau Gauge $\mathbf{A}=B_{y}z\mathbf{e}_{x}+B_{z}x\mathbf{e}_{y}$ where the vector potential doesn't have any component along the z-axis motion. This Hamiltonian can be expanded in the powers of $B_y$: to the lowest order we have ${\hat H} = {\hat H}_a + {\hat H}_l$, the first order in $B_y$ introduces an atom-cavity interaction term ${\hat H}_c = -e B_y z {\hat p}_x/m$. Writing ${\hat H}_c$ in terms of the Landau level creation and annihilation operators we obtain the following expression:

\begin{align}
\hat{\mathcal{H}}= \hbar \omega_c {\hat a}^+ {\hat a} + \sum_\alpha \varepsilon_{\alpha}\left|\alpha\right\rangle \left\langle \alpha\right| + \frac{\hbar \omega_y}{\sqrt{2}} \left(\hat{a}^{+}+\hat{a}\right) \frac{ {\hat z} }{ \ell_B } 
\label{totH}
\end{align}

\noindent In this equation we introduced the cyclotron frequency along the in plane field $\omega_y = e B_y/ m$ as well as the magnetic length for the perpendicular field $\ell_B = \sqrt{\hbar/(m \omega_c)}$, the notation ${\hat z}$ stands for the matrix elements of the $z$ operator on the vertical eigenstates $|\alpha\rangle$, it plays here the role of the dipole moment operator in quantum electrodynamics. The QED Hamiltonian Eq.~(\ref{totH}) appears in models where a photon mode/harmonic oscillator (Landau levels in our experiment) is coupled to an atom/qubit provided by Rydberg states. The strength of the interaction, which would be the vacuum Rabi splitting in atomic physics, is directly controllable and proportional to $B_y$ allowing in principle couplings of arbitrary strength.

This Hamiltonian may seem valid to first order in $B_y$, however the second order diamagnetic term $m \omega_y^2 z^2/2$ only renormalizes the vertical confinement potential $V_a(z) = V^0_a(z) + m \omega_y^2 z^2/2$. Thus Eq.~(\ref{totH}) remains valid for arbitrary interaction strength keeping in mind that the in-plane magnetic field then not only controls the coupling strength between the atom and Landau levels but also changes the atom energies $\epsilon_\alpha$ and the dipole momentum matrix ${\hat z}$, which can still be obtained easily by solving the one dimensional Schr\"odinger equation Eq.~(\ref{atom_energy}) in the modified confinement potential.

\begin{figure}[!htb]
\centering
	\includegraphics[width=1\columnwidth]{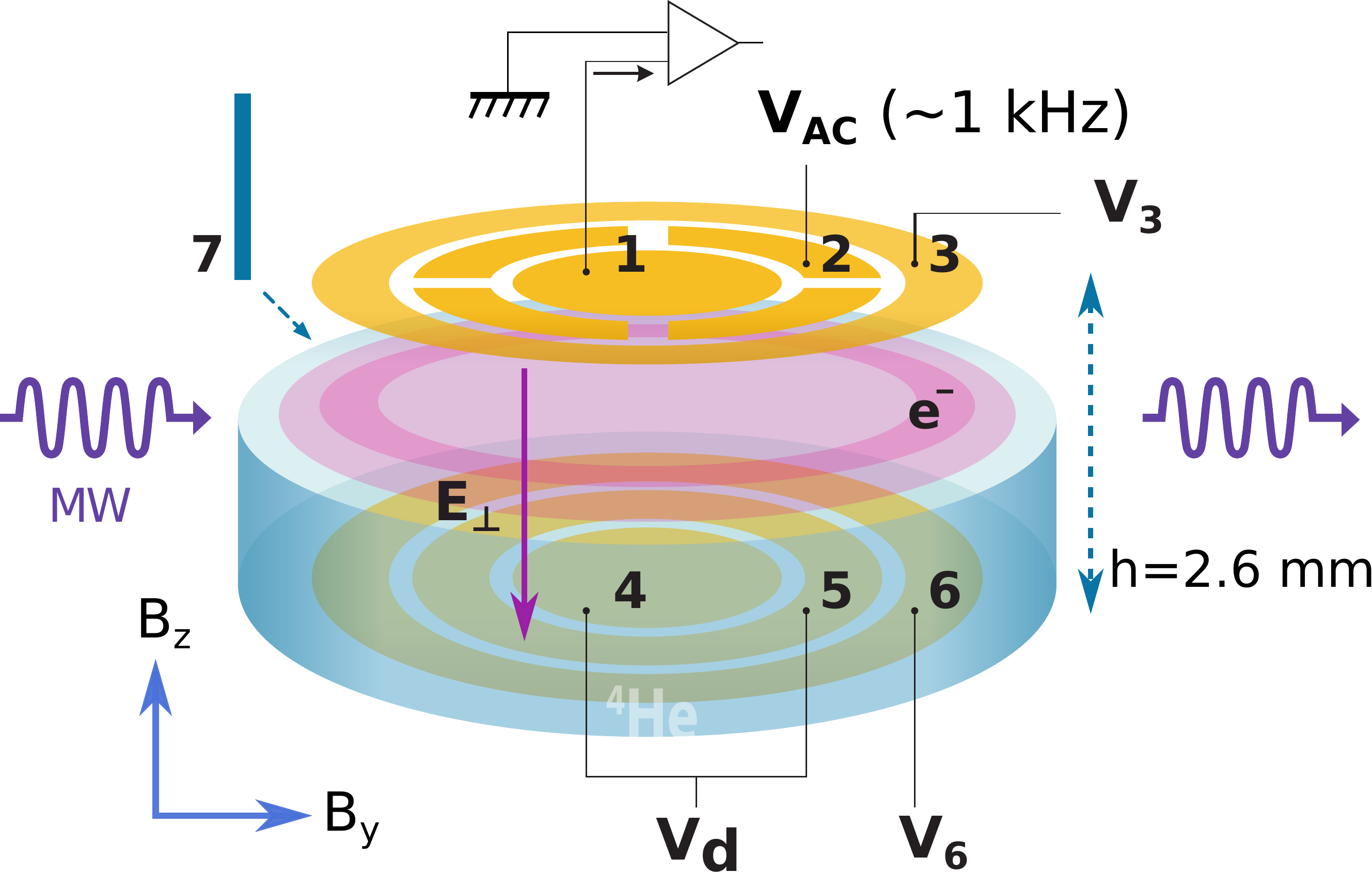}
	\caption{\label{cell} 
	Schematic diagram of the experimental cell. The top electrodes \textbf{\textit{1}} and \textbf{\textit{2}} are DC grounded and are used for the AC measurements. A positive DC voltage $V_d$ is applied to \textbf{\textit{4}} and \textbf{\textit{5}} confining the electrons into the center of the cell and fixing $E_\perp = V_d / h$. The electrodes \textbf{\textit{3}} and \textbf{\textit{6}} are used as a guard with negative potential. To ensure an homogeneous $E_\perp$ we fixed $V_{6} - V_{3}$ to $V_d$ (and $V_{d}-V_6 = 6{\rm V}$). The admittance $Y$ of the cell is obtained by applying a $10\;{\rm mV}$ AC voltage at $1137\;{\rm Hz}$ to the segmented electrode \textbf{\textit{2}} and measuring the induced pickup voltage on  \textbf{\textit{1}} with a lock-in amplifier. It depends on the in-plane conductivity of the electrons under magnetic field as obtained from Corbino measurements with Ohmic contacts on conventional 2DES. MW power was sent into the cell through a waveguide and  \textbf{\textit{7}} is the filament ($e^-$ source).
	}
\end{figure}

To check if QED Hamiltonian quantitatively describes the energy of the transitions between Rydberg states, we perform Stark effect spectroscopy. Electrons on helium form a static dipole with their image charge and the Stark effect due to the perpendicular electric field $E_\perp$ leads to a linear displacement of the Rydberg energy levels which can bring these atomic transitions in resonance with the external millimeter-wave excitation. At resonance a change of resistivity occurs due to MIRO allowing to detect the position of the energy levels. Our experimental setup consists of a cavity with Corbino electrodes, a layout is shown in Fig.~\ref{cell}. This cavity is half-filled with liquid helium by condensing helium vapor and monitoring the capacitance between top and bottom electrodes. Electrons are then deposited by thermal emission from a heated filament and are trapped on the surface by the pressing electric field $E_\perp$. They form a 2DES which behaves, for in plane transport, as an effective resistance $R$ placed between two contacts with capacitance $C$. This resistance can then be determined by measuring the admittance of the cell $Y$ between the two inner Corbino contacts from the top electrodes at frequencies comparable with the $R C$ relaxation time (we used $1137\;{\rm Hz}$). To extract the MW dependent (MIRO) admittance $\delta Y$, MW power is modulated at a frequency of $17\;{\rm Hz}$ and a double demodulation technique is used. \textcolor{black}{ Real and imaginary parts of $\delta Y$ give very similar lineshapes}. In our measurements the electron gas density was $n_e \simeq 1.5\times 10^7\;{\rm cm^{-2}}$ with a total number of $4\times 10^7$ electrons trapped in the cloud.

\begin{figure}[!htb]
\centering
	\includegraphics[width=0.8 \columnwidth]{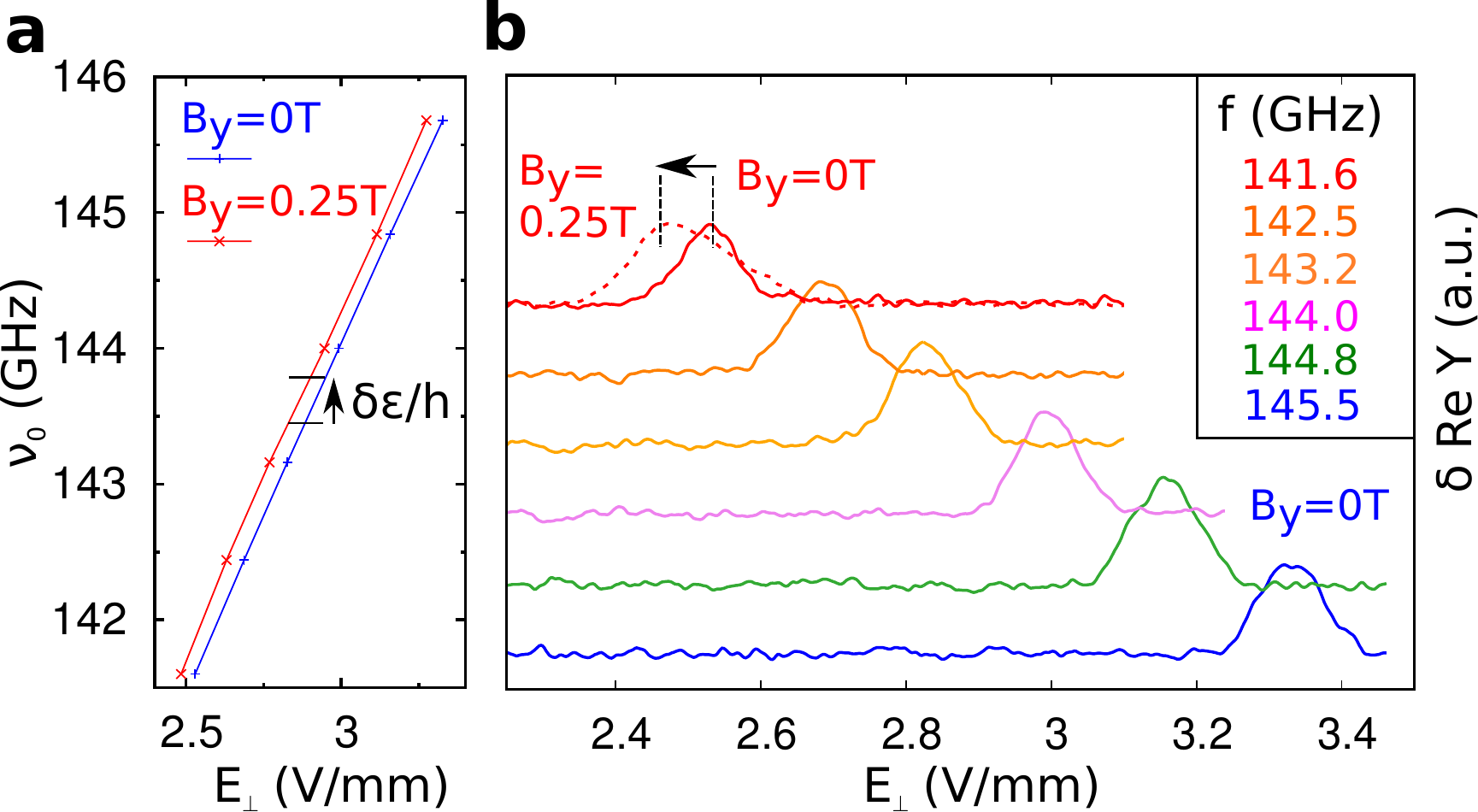}
	\caption{\label{v-calib}  
	The panel {\bf a} shows the shift of the $|g\rangle \rightarrow |e\rangle$ transition energy due to Stark effect at $B_{z} = 0.73\;{\rm T}$ for $B_{y} = 0$ and $B_{y}=0.25\;{\rm T}$. The slope is the same for both lines but a small energy shift $\delta \epsilon$ is observed. The panel {\bf b} illustrates the Stark shift as seen from raw ${\delta Y}$ data at $B_{y} = 0$. The dashed line displays the shift of the resonance with $B_y$ for $f = 141.6\;{\rm GHz}$, $\delta \epsilon$ can also be deduced from the value of this Stark shift.
	}
\end{figure}

The conversion between Stark shifts and transition energies is obtained from a calibration experiment where we excite the electrons with photons at different energies and measure the electric field at resonance (see Fig.~\ref{v-calib}). For weak parallel magnetic fields the transition from the ground $|g\rangle$ to the first excited Rydberg state $|e\rangle$ manifests as a resonance of the microwave induced change in admittance as function of $E_\perp$.
The resonance position at energy $h \nu_0 = \epsilon_e - \epsilon_g$  shifts linearly with  $E_\perp$, the slope can be obtained from the Schr\"odinger equation  Eq.~(\ref{atom_energy}) with small deviations due to uncertainties on geometrical parameters.
This slope is almost independent of $B_y$ (see Fig.~\ref{v-calib}.{\bf a}), indeed for $B_y \le 1~{\rm Tesla}$ the coupling term of the QED Hamiltonian $\frac{\hbar \omega_y}{\sqrt{2}} \frac{ \langle e|{\hat z}|g\rangle }{ \ell_B } \lesssim 10\;{\rm GHz}$ is small compared to $h \nu_0 \simeq 140$  GHz and does not change the vertical dipole moment significantly. While the slope as function of $E_\perp$ remains unchanged, an overall energy shift $\delta \epsilon$ is visible. \textcolor{black}{
It appears due to the coupling between Rydberg states and Landau levels at finite $B_y$. In the following, we present a careful experimental investigation of the coupling induced energy shift and show that it can be understood quantitatively from the QED Hamiltonian.
}

\FloatBarrier

\begin{figure*}[!htb]
\centering
 \includegraphics[width=0.8\paperwidth, height =0.5\paperwidth ]{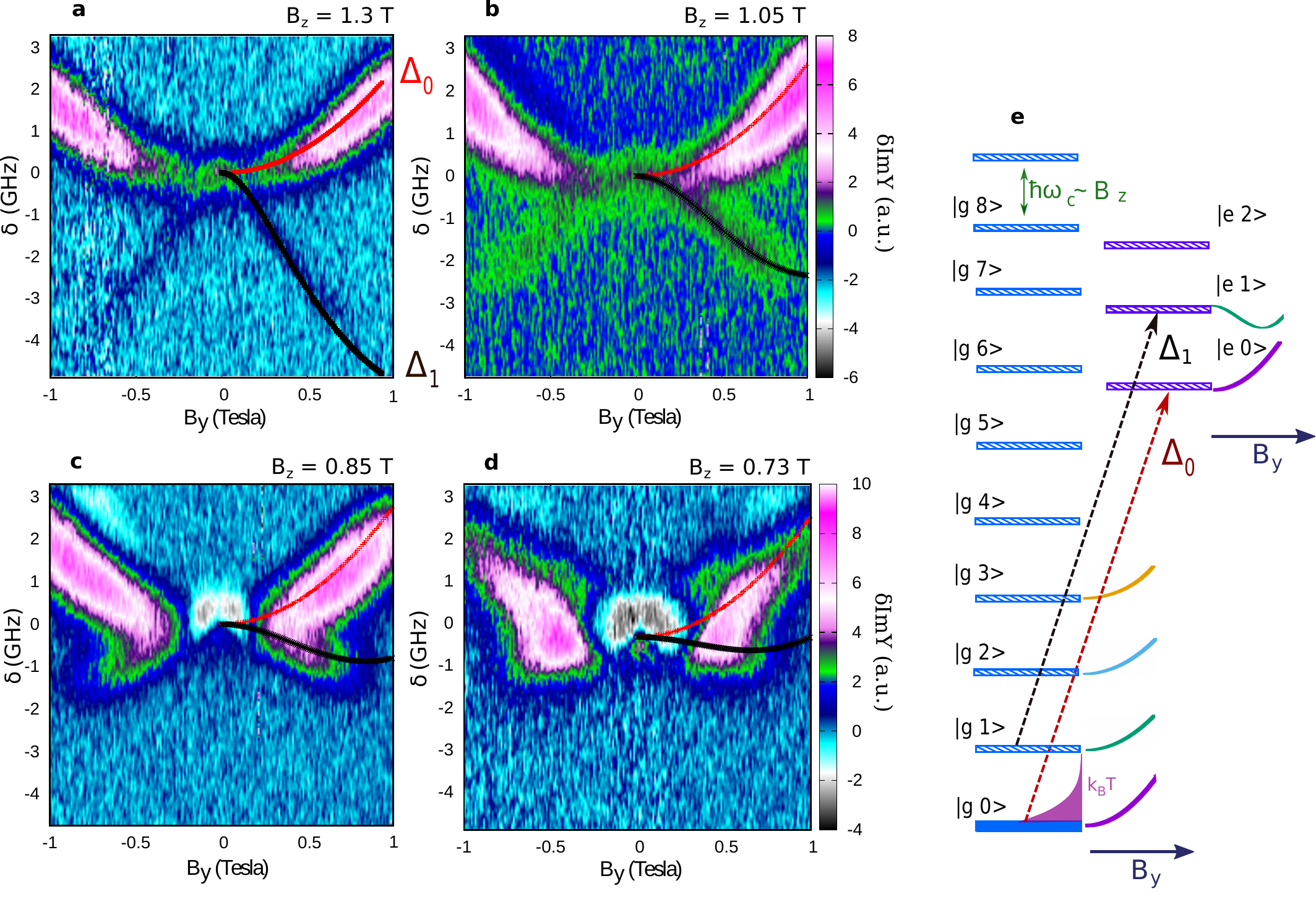}
 \caption{\label{pic-map} Evolution of the $|g\rangle \rightarrow |e\rangle$ resonance with $B_y$ showing that this resonance splits into two branches. The colour maps represent $\delta Y$  as function of $B_y$ and detuning $\delta$ for $B_{z} = $ 1.3 (\textbf{a}), 1.05 (\textbf{b}), 0.85 (\textbf{c}) and 0.73 T (\textbf{d}). Red and black curves give the QED Hamiltonian predictions for the $\Delta_0$ and $\Delta_1$ transitions drawn in panel (\textbf{e}) between states $|n,m\rangle$ where the first quantum number gives the atomic state and $m$ is Landau level number (number of photons in the cavity). The  calculated evolution of individual levels (rescaled for visibility) with $B_y$ up to 1 T is shown by coloured lines.
 }
\end{figure*}

To study the evolution of the $|g\rangle \rightarrow |e\rangle$ transition with $B_y$, we take advantage of the linear dependence of the energy shifts on $E_\perp$, which enables us to fix the excitation frequency to $f = 139\;{\rm GHz}$ and change only $E_\perp$. 
\textcolor{black}{
We define $\delta$ as the detuning induced by the Shark shift due to the deviation of $E_\perp$ from its resonant value at the excitation frequency $f$ for $B_y = 0$; $\delta$ is thus the difference between $E_\perp$ and its value at resonance $\simeq 2\;{\rm V mm^{-1}}$ times (minus) the slope measured in Fig.~\ref{v-calib}.{\bf a}.}
All the collected data is then transformed into a map where the change in admittance is plotted as function of $B_y$ and of the detuning $\delta$. 
The maps that we obtained for $B_z = 1.3, 1.05$ are displayed on Fig.~\ref{pic-map}(a,b). When increasing $B_y$ we can resolve two transitions $\Delta_0$ and $\Delta_1$. The energy of the more intense transition called $\Delta_0$ increases with $B_y$, quadratically at weak $B_y$ with a cross-over to a more linear dependence at the highest fields. In addition, to this main transition a weaker transition $\Delta_1$ splits off from the main transition as $B_y$ becomes stronger (as will be shown later $\Delta_1$ is visible only at highest microwave excitation power which was here fixed to its maximal value of 10 mW). This  $\Delta_1$ transition becomes as visible as the main transition at lower $B_z$ (Fig.~\ref{pic-map}(c,d) for $B_z = 0.85, 0.73$~T) giving two mutually inverse curves with a characteristic "butterfly" pattern. The coupling strength dependence of $\Delta_0$ is almost the same for all $B_z$ in our data set, on the contrary for $\Delta_1$ the slope of the transition line as function of the coupling strength increases significantly with $B_z$.

The splitting of the Rydberg transition can be understood from the energy level diagram in Fig.~\ref{pic-map}(e) which shows how the energy levels from the QED Hamiltonian evolve with the coupling strength. For each atomic state $|\alpha\rangle$, the manifold of dressed states consists of a ladder of Landau levels $|\alpha,m\rangle$ with MW exciting transitions that conserve the Landau level number $m$ (between states with the same number of photons in the cavity). \textcolor{black}{ Without parallel magnetic field the energy of the $|g,m\rangle \rightarrow |e,m\rangle$ transition does not dependent on $m$. The coupling lifts this degeneracy, making transitions associated to different Landau levels spectroscopically distinguishable. In the special case of the $m = 0$ transition $|g, 0\rangle \rightarrow |e,0\rangle$ with energy $\Delta_0$ the renormalization of the transition energy is due to an interaction with the lowest Landau level. It can be seen as an effective Lamb shift in analogy with atomic physics, circuit QED \cite{qed_lamb,circuit_qed_qubit} and physics of electrons coupled to phonons or ripplons \cite{qdot,ripplon_lamb}.} A similar renormalization occurs for all the transitions $|g,m\rangle \rightarrow |e,m\rangle$, and simulations are thus needed to identify the observed spectroscopic lines as one of the transitions $\Delta_m$.

The QED Hamiltonian gives a quantitative prediction on the renormalization of the transition energies $\Delta_m$, we emphasize that all the parameters appearing in the model involve $E_\perp$, the applied magnetic field, the liquid Helium dielectric constant $\epsilon$ and fundamental constants, there are thus no fitting parameters. The values of $\Delta_m$ can be obtained from the numerical diagonalization of Eq.~(\ref{totH}), to obtain accurate values we had to include Rydberg states and Landau levels at an energy scale higher than $\hbar \nu_0$ from $|g,0>$, in the simulations shown here we used a basis set of $100$ Landau levels and $20$ Rydberg states. 
Results of our simulations for transitions $\Delta_{0,1}$ are overlaid on top of the experimental data. We see that they reproduce accurately both upper and lower "butterfly wings", including the striking increase of $\Delta_1(B_y)$ with $B_z$ which contrasts with the $\Delta_0$ transition that is almost $B_z$ independent.

\begin{figure}[!htb]
\centering
 \includegraphics[width=1\columnwidth]{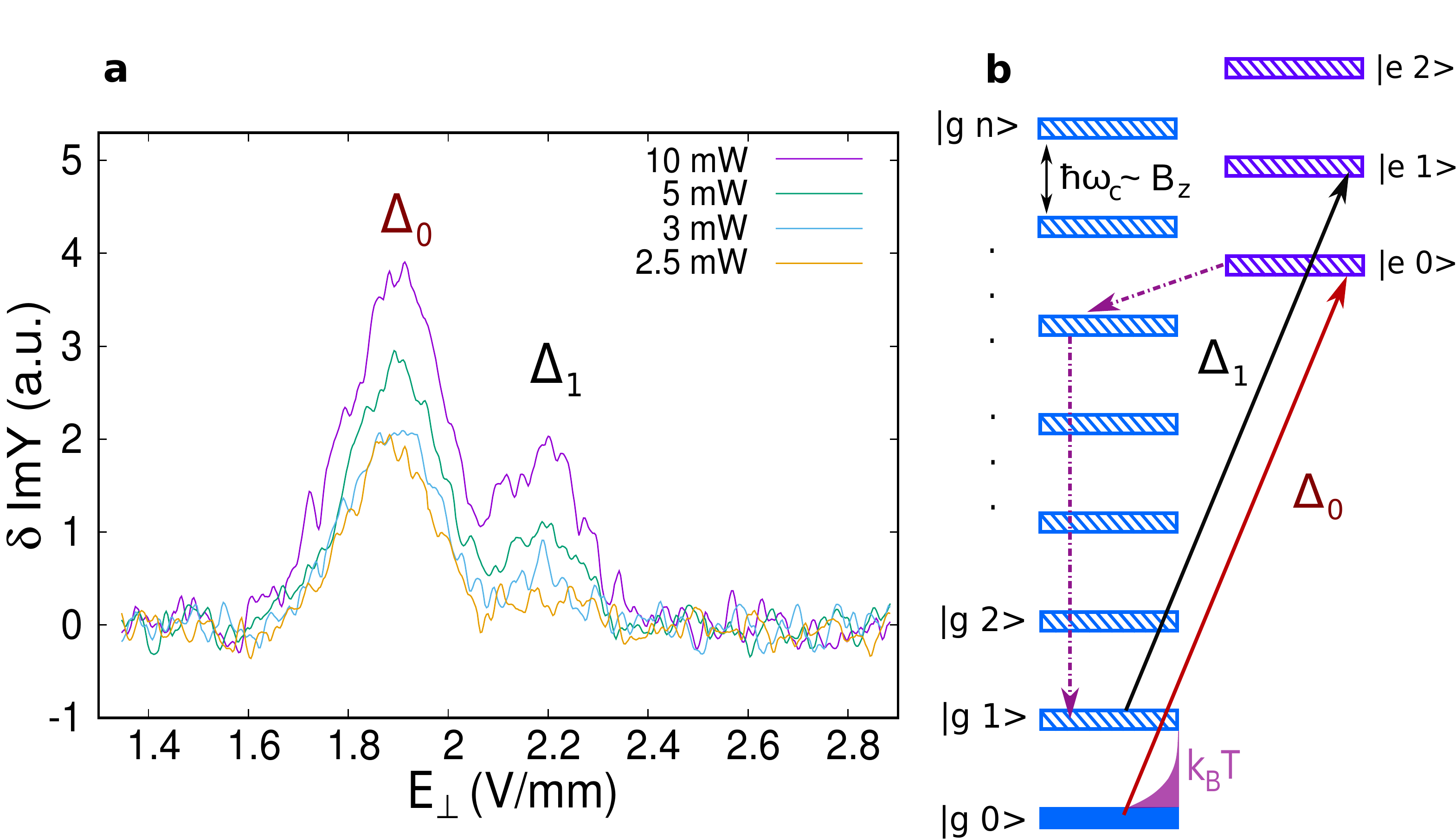}
 \caption{\label{pop-n-power} \textbf{a} Dependence of the $\Delta_{0,1}$ resonances on MW excitation power at $B_{z}$ = 1 T, $B_{y}$ = 0.5 T. The $\Delta_1$ transition disappears at low power. \textbf{b} illustrates how the $\Delta_0$ transition can populate the $|g,1\rangle$ level, if the energies  $\Delta_{0,1}$ are close, MW photons at the $\Delta_1$ energy can also induce the $\Delta_0$ transition providing the non-equilibrium population needed to see the resonance at $\Delta_1$.
 }
\end{figure}

The transitions $\Delta_m$ between states $|g,m\rangle \rightarrow |e,m\rangle$ can only be observed if the initial state $|g,m\rangle$ is populated. At the experiment temperature 
$T = 0.3\;{\rm K}$ only the ground state $|g,0\rangle$ is populated in equilibrium (the thermal population of $|g, 1\rangle$ at T=0.3~K and $B_z = 1\;{\rm T}$ is only 1\%). As a consequence, the transitions $\Delta_m$ with $m \ge 1$ require an external excitation to become visible.
In Fig.~\ref{pop-n-power}(a) we show that indeed these transitions physically appear only when the MW power is high enough, as opposed to the $\Delta_0$ transition which is present even at low MW power. Two possible mechanisms to populate the $|g,1\rangle$ level may be taken into consideration. The first assumes in plane component of the MW electric field populating a $|g,1\rangle$ level non-resonantly from the initial $|g,0\rangle$ level. The second one is illustrated with dashed lines on the Fig.~\ref{pop-n-power}(b). When the energy of the transitions $\Delta_{0,1}$ are sufficiently close (at low $B_y$) a $\Delta_1$ energy MW photon can also excite a transition into the $|e,0\rangle$ state, \textcolor{black}{scattering can then transfer some population into a nearby $|g,m\rangle$ level, leading (after relaxation) to a finite population in the $|g,1\rangle$ state which makes the transition $\Delta_1$ visible}. In Fig.~\ref{pic-map} we see that as $B_y$ increases the $\Delta_1$ transition disappears faster than the $\Delta_0$ transition. 
This observation can be understood within the population mechanism for $|g,1\rangle$ shown in Fig.(\ref{pop-n-power}b). Indeed at larger $B_{y}$ the energies of the $\Delta_{0,1}$ transitions become different and the MW excitation at the $\Delta_1$ frequency can no longer excite the $\Delta_0$ transition which populates the $|g,1\rangle$ level. The state $|g,1\rangle$ then remains empty leading to the disappearance of the $\Delta_1$ wings. 

In conclusion we have shown that 2DES on liquid helium can form a prototype quantum system of an atom coupled to an oscillator with the coupling directly controllable by the parallel magnetic field. Our spectroscopic results compare very accurately to theoretical predictions with no adjustable parameters. Control of the population transfer between dressed states could enable tunable mm-wave lasers and future experiments at high electron densities may reveal a rich quantum many body physics.


We acknowledge discussions with K. Kono, D.L. Shepelyansky who stimulated these experiments as well as support from N. Vernier, V.P. Dvornichenko, Okinawa Institute of Science and Technology (OIST) Graduate University and ANR SPINEX. 

\bibliography{REFs2DEG.bib}

\end{document}